\documentclass{amsart}
 \usepackage{amssymb,enumerate,gacspf,gacs}

\numberwithin{equation}{section} 
\numberwithin{figure}{section} 

\makeatletter
\newcommand\amod{\mskip-\medmuskip\mkern5mu
  \mathbin{\operator@font amod}\penalty900\mkern5mu\mskip-\medmuskip}
\makeatother

\renewcommand{\AA}{\mathbb{A}}
\newcommand{\f}{f}
\newcommand{\F}{\var{F}}
\newcommand{\G}{\var{G}}
\newcommand{\Cur}{\var{Cur}}
\newcommand{\Prev}{\var{Prev}}
\newcommand{\Age}{\var{Age}}

\pfshortnumbers{4} 

\begin{document}

  \title[Deterministic computations]{Deterministic
          computations
 \\  whose history is independent
 \\ of the order of asynchronous updating}

\author{Peter G\'acs}
\address{Computer Science Department\\ Boston University}
\email{gacs@bu.edu}

\date{\today}

\thanks{Partially supported by NSF grant CCR-920484}

\begin{abstract}
  Consider a network of processors (sites) in which each site $x$ has
a finite set $N(x)$ of neighbors.
  There is a transition function $f$ that for each site $x$ computes the
next state $\xi(x)$ from the states in $N(x)$.
  But these transitions (updates) are applied in arbitrary order, one
or many at a time.
  If the state of site $x$ at time $t$ is $\eta(x,t)$ then let us
define the sequence $\zg(x,0),\zg(x,1)$, $\ldots$ by taking the
sequence $\eta(x,0),\eta(x,1)$, $\ldots$, and deleting each
repetition, i.e.~each element equal to the preceding one.
  The function $f$ is said to have \df{invariant histories} if the sequence
$\zg(x,i)$, (while it lasts, in case it is finite) depends only on the
initial configuration, not on the order of updates.

  This paper shows that though the invariant history property is typically
undecidable, there is a useful simple sufficient condition, called
\df{commutativity}:
  For any configuration, for any pair $x,y$ of neighbors, if the updating
would change both $\xi(x)$ and $\xi(y)$ then the result of updating first
$x$ and then $y$ is the same as the result of doing this in the reverse
order.
  This fact is derivable from known results on the confluence of
term-rewriting systems but the self-contained proof given here may be
justifiable.
 \end{abstract} 

\maketitle

\section{Introduction}\label{s.intro}

  Consider a set $\CC$ of processors (sites) in which each site $x$ has a
set $\SS$ of possible states (also called ``local states'').
  An arbitrary function $\xi\in\SS^{\CC}$ is called a
\df{space-configuration}, or simply ``configuration'', or ``global state''.
  The value $\xi(x)$ is the state of site $x$ in $\xi$.
  A function $N:\CC\to 2^{\CC}$ will be called a \df{neighborhood function}
assigning to each $x\in\CC$, a set $N(x)$ called the \df{neighborhood} of
$x$.
  A function $\f:\SS^{\CC}\to\SS^{\CC}$ is called a \df{transition
function} if $\f(\xi)(x)$ depends only on $\xi\on N(x)$, i.e.
 \[
  \xi_{1}\on N(x)=\xi_{2}\on N(x) \imp \f(\xi_{1})(x) = \f(\xi_{2})(x).
 \]
  The transition function determines a possible ``next'' configuration from
the ``current'' one.
  The 4-tuple
 \begin{equation}\label{e.autom}
  \AA = (\CC, \SS, N, \f)
 \end{equation}
  will be called an \df{automaton} (not necessarily a finite one).
  If all sets $N(x)$ are finite then the system is called \df{local}.
  Note that locality is actually a property of $\f$ itself: it says that
for each $x$ a finite $N(x)$ can be chosen such that $\f(\xi)(x)$ depends
only on $\xi\on N(x)$.
  Let $\ZZ_{+}=\ZZ\cap[0,\infty)$.

 \begin{example}[Cellular automata]\label{x.CA}\
 \begin{enumerate}

  \item On the set of integers:
  Let $\CC=\ZZ$, $N(x)=\{x-1,x,x+1\}$.
  Suppose that there is a transition function $g(x,y,z)$ and each site $x$
has a value $\xi(x)\in \SS$.
  Now the result of transition at site $x$ is 
 \[
   \f(\xi)(x) = g(\xi(x-1),\xi(x), \xi(x+1)).
 \]
  In this example, the transition function depends only on the sequence of
values of $\xi\on N(x)$, i.e.~it is homogenous.
  The present paper will not exploit any consequences of homogeneity.

  \item On the set of natural numbers, with ``free boundary condition'':
  Let $\CC=\ZZ_{+}$, $N(x)=\{x-1,x,x+1\}$ for $x>0$ and $\{0,1\}$ for $x=0$.
  Suppose that there are transition functions $g(x,y,z)$, $g_{0}(x,y)$.
  Now the result of transition at site $x$ is
 $g(\xi(x-1),\xi(x), \xi(x+1))$ for $x>0$ and $g_{0}(\xi(0), \xi(1))$ for
$x=0$.
 \end{enumerate}
 \end{example}

  Let us fix an automaton $\AA$ as in ~\eqref{e.autom}.
  An arbitrary function $\eta: \CC\times \ZZ_{+}\to \SS$ is called a
\df{space-time configuration}.
  Such a space-time configuration can also be viewed as a sequence $\eta:
\ZZ_{+}\to\SS^{\CC}$ of space-configurations.
  We will say that a space-time configuration $\eta$ is a \df{synchronous
trajectory} if for all $x,t$ we have $\eta(\cdot,t+1)=f(\eta(\cdot,t))$.
  In other words,
 \begin{equation}\label{e.update}
 \eta(x,t+1)=\f(\eta(\cdot,t))(x),
 \end{equation}
 i.e.~in $\eta$, each site is ``updated'' every time by the function $f$
(though the update may not change the state).
  We are interested in situations when at any one time, only the values of 
some of the sites are updated.
  We will say $\eta$ is an \df{asynchronous trajectory} if
~\eqref{e.update} holds for all $x,t$ such that
 $\eta(x,t+1)\neq \eta(x,t)$: i.e.~if each site in $\eta$ at each time is
either updated or left unchanged.
  From now on, when we speak of a ``trajectory'' without qualification,
this will mean an asynchronous trajectory.
  Let the \df{update set}
 \[
   U(t,\eta)
 \]
  be the set of sites $x$ with $\eta(x,t+1)\neq \eta(x,t)$.
  The initial configuration and the update sets $U(t,\eta)$ determine
$\eta$.
  For any set $A$, let
 \[
   \chi(x,A)=\begin{cases}
   1 &\text{if $x\in A$,}
\\ 0 &\text{otherwise.}
	     \end{cases}
 \]
  For a space-time configuration $\eta(x,t)$ we define the function
$\tau(x,t)=\tau(x,t,\eta)$ as follows:
 \begin{align*}
	\tau(x,0)   &= 0,
\\ 	\tau(x,t+1) &= \tau(x,t)+ \chi(x,U(t,\eta)).
 \end{align*}
  We can call $\tau(x,t)$ the \df{effective age} of site $x$ in the
space-time configuration $\eta$ at time $t$: this is the number of effective
updatings that $x$ underwent until time $t$.
  Given an initial configuration $\xi$, we say that $\f$ (and thus $\AA$)
has \df{invariant histories} on $\xi$ if there is a function
$\zg(x,u)=\zg(x,u,\xi)$ such that for all asynchronous trajectories
$\eta(x,t)$ with $\eta(\cdot,0)=\xi$ we have
 \begin{equation}\label{e.t.commut}
   \eta(x,t) = \zg(x,\tau(x,t,\eta),\xi).
 \end{equation}
  This means that after eliminating repetitions, the sequence
$\zg(x,0),\zg(x,1)$, $\ldots$ of values that a site $x$ will go through
during some space-time configuration, does not depend on the update sets,
only on the initial configuration (except that the sequence may be finite
if there is not an infinite number of successful updates).
  The update sets influence only the delays in going through this
sequence.
  We say that an automaton has \df{invariant histories} if it has such on
all initial configurations.

 \begin{remark}
  The sequence $\zg(x,0),\zg(x,1),\ldots$ is a sequence of local states but
$\zg(\cdot,n)$ is not a space-configuration (global state) that appears at
any time in a typical asynchronous trajectory.
 \end{remark}

  \begin{theorem}\label{t.undec}
  If $\AA$ is a one-dimensional cellular automaton with state space
$\SS=\{0,\ldots,n-1\}$ for some natural number $n$, then it is undecidable
whether it has invariant histories.
  \end{theorem}

  The theorem shows that some more condition is needed if we want the
invariant history property to become decidable.
  For us, this condition will be monotonicity.
  The set of \df{free} sites $x$ in a configuration $\xi$ is defined by
 \[
    L(\xi)=\setof{x : \f(\xi)(x)\neq \xi(x)}.
 \]
  For a space-time configuration $\eta$, let
 \[
  L(t,\eta)=L(\eta(\cdot,t)).
 \]
  For a configuration $\xi$ and a set $E$ of sites, let 
 \begin{align*}
  \f(\xi,E)(x) &= \begin{cases}
  \f(\xi)(x) 	&\text{if $x\in E$}
\\\xi(x) 	&\text{otherwise.}
                 \end{cases}
\\ \f(\xi,E,F) &=\f(\f(\xi,E),F).
 \end{align*}
  With this notation, we have $\f(\xi) = \f(\xi,\CC) = \f(\xi, L(\xi))$.
  Now we can express the condition that $\eta$ is an asynchronous
trajectory by saying that for every $t$ there is a set $U$ with
 \begin{equation}\label{e.asynch.traj}
 \eta(\cdot,t+1)=\f(\eta(\cdot,t),U),
 \end{equation}
  and the condition that $\eta$ is synchronous by requiring
$U(t,\eta)=L(t,\eta)$ for each $t$.
  We call a transition rule $\f$ \df{monotonic} if
 $L(t,\eta)\xcpt U(t,\eta)\sbsq L(t+1,\eta)$,
  i.e.~updating a site cannot take away the freedom of other sites.
  We call a transition rule $\f$ (and thus the automaton $\AA$)
\df{commutative} if for all configurations $\xi$ and all disjoint sets of
sites $A,B\sbsq L(\xi)$ we have
 \begin{equation}\label{e.commut.gener}
   \f(\xi,A,B) = \f(\xi,A\cup B).
 \end{equation}
  We call $\f$ \df{locally commutative} when this property is required just
for the special case where $A,B$ are one-element sets.
  The following fact is easy to see but we give the proof for completeness.

  \begin{lemma}\label{l.commut.gener}
  If $\f$ is local then its local commutativity implies commutativity.
  \end{lemma}
  \begin{proof}
  Let us first show
 \begin{equation}\label{e.commut.gener.1}
  \f(\xi,\{x_{1}\},\dots,\{x_{n}\})=\f(\xi,\{x_{1},\ldots,x_{n}\}).
 \end{equation}
  Local commutativity implies for each $k$,
 \[
   \xi'=\f(\xi,\{x_{1}\},\ldots,\{x_{n}\}) =
  \f(\xi,\{x_{k}\},\{x_{1}\},\dots,\{x_{k-1}\},\{x_{k+1}\},
    \ldots,\{x_{n}\}).
 \]
  Therefore $\xi'(x_{k})=\f(\xi,\{x_{1},\dots,x_{n}\})(x_{k})$.
  Now, let us show
 \begin{equation}\label{e.commut.gener.2}
   \f(\xi,\{x_{1},\dots,x_{n}\},\{y\}) = \f(\xi,\{x_{1},\dots,x_{n},y\}).
 \end{equation}
  Using ~\eqref{e.commut.gener.1}, we have
 $\f(\xi,\{x_{1}\},\dots,\{x_{n}\})=\f(\xi,\{x_{1},\ldots,x_{n}\})$,
 hence
 $\f(\xi,\{x_{1},\ldots,x_{n}\},\{y\})=
  \f(\xi,\{x_{1}\},\ldots,\{x_{n}\},\{y\})$.
  Using ~\eqref{e.commut.gener.1} again concludes the proof.
  
  Let us return to the general case.
  Obviously, it is sufficient to check ~\eqref{e.commut.gener} for sites
$y\in B$.
  Clearly, $\f(\xi,A,B)(y) = \f(\xi,N(y)\cap A,\{y\})$.
  The latter is $\f(\xi,(N(y)\cap A)\cup\{y\})$ according to
~\eqref{e.commut.gener.2}.
 \end{proof}

 \begin{remarks}\
 \begin{enumerate}

  \item
  For the cellular automaton example above, local commutativity is
equivalent to saying that if $g(r_{0},r_{1},r_{2})\neq r_{1}$ and
$g(r_{1},r_{2},r_3)\neq r_{2}$ then
 \begin{align*}
   g(g(r_{0},r_{1},r_{2}),r_{2},r_3) &= g(r_{1},r_{2},r_3)
\\ g(r_{0},r_{1},g(r_{1},r_{2},r_3)) &= g(r_{0},r_{1},r_{2}).
 \end{align*}
  
  \item
  If $\f$ is not local then local commutativity does not always imply
commutativity.
  For an example, let $\CC=\{0,1\}$, $\CC=\ZZ$, $N(x)=\CC$, and let 
 \[
  \f(\xi)(x)=\begin{cases}
   1 &\text{if $\xi(y)=0$ for all but finitely many $y$,}
\\ 0 &\text{otherwise.}
	     \end{cases}
 \]
  Now $\f$ is obviously locally commutative.
  On the other hand, let $\xi_{0}(x)=0$ for all $x$, and let
  Then $\f(\xi_{0},\ZZ)(-1)=1$ and
 $\f(\xi_{0},\ZZ_{+},\ZZ\xcpt\ZZ_{+})(-1)=0$.
 \end{enumerate}
 \end{remarks}

 \begin{theorem}\label{t.commut}
  A transition function is commutative if and only if it is monotonic and
has invariant histories.
 \end{theorem}

  In Theorem~ \ref{t.asynch-sim} below, we will give a known simple
example of a commutative transition function.
  For that example, the theorem can be proved much easier.

  Theorem \ref{t.commut} can be derived from results e.g. in
~\cite{HuetConflu80}.
  However, I do not find it worth introducing all the concepts needed for
the derivation: the simplicity of the condition in the present context
probably justifies a self-contained proof.

  \section{Commutativity implies invariant histories}

  \begin{lemma}\label{l.commut.necess}
  Suppose that $\f$ has invariant histories and is monotonic: then it is
commutative.
  \end{lemma}
  \begin{proof}
  Let $U_{1}(0)=U_{2}(1)=\{x\}$, $U_{1}(1)=U_{2}(0)=\{y\}$, and
$U_{1}(t+2)=U_{2}(t+2)$.
  This defines $\eta_{1}$ and $\eta_{2}$ from initial configuration $\xi$
by $U_{1},U_{2}$ as usual.
  By monotonicity, $\eta_{1}(y,1)\neq\eta_{1}(y,2)$ and
$\eta_{2}(x,1)\neq\eta_{2}(x,2)$, so $\tau$'s values satisfy
 \[
   \tau(x,2,\eta_{1}) = \sum_{t=0}^{1}\chi(w, U_{1}(t,\eta_{1}))
 \]
  which is 1 if $w\in\{x,y\}$ and 0 otherwise.
  The same value is obtained for $\tau(x,2,\eta_{2})$.
  By invariant histories, there is a $\zg$ such that 
 \[
   \eta_{1}(w,2)=\zg(w,\tau(w,2,\eta_{1}))
  = \zg(w,\tau(w,2,\eta_{2})) = \eta_{2}(w,2)
 \]
  and
 \[
  f(\xi,{x},{y}) = f(\xi,U_{1}(0),U_{1}(1)) = \eta_{1}(w,2)
   = \eta_{2}(w,2) = f(\xi,U_{2}(0),U_{2}(1)) = f(\xi,\{y\},\{x\}).
 \]
  Thus, $f$ is commutative.
\end{proof}

  What remains to prove after Lemma~\ref{l.commut.necess} is that
commutativity implies monotonicity and invariant histories.

  \begin{lemma}\label{l.commut.monot}
  If $\f$ is commutative then it is monotonic.
  \end{lemma}
  \begin{proof}
  By Lemma~\ref{l.commut.gener}, 
 $f(\xi, U(t,\eta), L(t,\eta) \xcpt U(t,\eta)) = f(\xi, L(r,\eta))$.
  Therefore
 $L(t,\eta) \xcpt U(t,\eta)\sbsq L(t,\eta)$ implies that $f$ is monotonic.
 \end{proof}

  We say for two asynchronous trajectories $\eta_{0},\eta_{1}$ with the
same initial configuration that $\eta_{1}$ \df{dominates} $\eta_{0}$ until
time $u$ if the following conditions hold:
 \begin{cjenum}
  \item\label{i.domin.ineq}
   $\tau(\cdot,t,\eta_{0})\le\tau(\cdot,t,\eta_{1})$ for all $t\le u$
  \item\label{i.domin.eq}
   for all $t_{0},t_{1}\le u$, if
 $\tau(x,t_{0},\eta_{0})=\tau(x,t_{1},\eta_{1})$ then
 $\eta_{0}(x,t_{0})=\eta_{1}(x,t_{1})$.
 \end{cjenum}
  When $\eta_{1}$ dominates $\eta_{0}$ up to time $u$ for all $u$ then we
simply say that $\eta_{1}$ dominates $\eta_{0}$.
  This domination is, of course, a transitive relation.
  If the rule has invariant histories then condition ~\eqref{i.domin.ineq}
implies ~\eqref{i.domin.eq}, but otherwise this may not be the case.

 \begin{Proof}[Proof of Theorem~\protect\ref{t.commut}]
  Let $\f$ be a commutative transition rule.
  It remains to prove that it has invariant histories.
  
  \begin{step+}{commut.main-1}
  Let $\eta$ be an asynchronous trajectory and
 $A_{0}\sbsq L(0,\eta)\xcpt U(0,\eta)$.
  Then there is an asynchronous trajectory $\eta'$ dominating $\eta$ with
initial configuration $\eta(\cdot,0)$, such that
 $U(0,\eta')=U(0,\eta)\cup A_{0}$.
  \end{step+}
  \begin{pproof}
  Let $\xi_{0}=\eta(\cdot,0)$.
  We show how to build, for each $u$, a trajectory $\eta'$ with the given
properties that dominates $\eta$ up to time $u$.
  When $u\to\infty$ then $\eta'$ will converge to a trajectory with the
same properties that dominates $\eta$.
  For $u=0$ we can choose $\eta'(\cdot,0)=\eta(\cdot,0)$.
  We assume that $\eta'$ can be constructed for all $v<u$ and prove it for
$u$.
  Let $\xi_{1}=\eta(\cdot,1)$, and $A_{1}=A_{0}\xcpt U(1,\eta)$.
  Let the trajectory $\eta_{1}$ be defined by $\eta_{1}(x,t)=\eta(x,t+1)$.
  The inductive assumption gives a trajectory $\eta'_{1}$ with initial
configuration $\xi_{1}$ dominating $\eta_{1}$, with
 \begin{equation}\label{e.U}
  U(0,\eta'_{1}) = A_{1}\cup U(0,\eta_{1}).
 \end{equation}
  Using this trajectory, we define, for $t>0$:
 \[
   \eta'(\cdot,t)=\begin{cases}
     \f(\xi_{0},A_{0}\cup U(0,\eta)) 	&\text{if $t=1$,}
\\   \eta'_{1}(\cdot,t-1)		&\text{otherwise.}
	      \end{cases}
 \]
 \begin{step+}{commut.main-1.traj}
  $\eta'$ is an asynchronous trajectory.
 \end{step+}
 \begin{pproof}
  Let us show that $\eta'$ satisfies ~\eqref{e.asynch.traj}.
  This holds by definition for $t=0$ and $t>1$.
  Let us show that it also holds for $t=1$ with $U=U(1,\eta)\xcpt A_{0}$.
  We have 
 \begin{equation}\label{e.commut.eta-pr-1}
 \begin{aligned}						       
   \eta'(\cdot,2) &= \eta'_{1}(\cdot,1)
   &&\text{ by def.},
\\ &= \f(\xi_{1}, A_{1}\cup U(0,\eta_{1}))
   &&\text{ by~\eqref{e.U}},
\\ &= \f(\xi_{1}, A_{1}\cup U(1,\eta))
   &&\text{ by def.~of } \eta_{1},
\\ &= \f(\xi_{1}, (A_{0}\xcpt U(1,\eta)) \cup U(1,\eta))
   &&\text{ by def.~of $A_{1}$},
\\ &= \f(\xi_{1}, A_{0}\cup (U(1,\eta)\xcpt A_{0}))
\\ &= \f(\xi_{0}, U(0,\eta), A_{0}, U(1,\eta)\xcpt A_{0})
   &&\text{ by def.~of $\xi_{1}$ and commutativity},
\\ &= \f(\eta'(\cdot,1), U(1,\eta)\xcpt A_{0}).
\end{aligned}
 \end{equation}
\end{pproof}

  For domination, we must check two properties.
  
 \begin{step+}{commut.main-1.tau}
  We have  $\tau(x,t,\eta)\le\tau(x,t,\eta')$.
 \end{step+}
 \begin{pproof}
  By the definition of $\tau$, for $t>0$,
 \[
  \tau(x,t,\eta)= \begin{cases}
    \chi(x,U(0,\eta))  				&\text{if $t=1$,}
\\  \tau(x,1,\eta) + \tau(x,t-1,\eta_{1}) 	&\text{if $t>1$.}
                  \end{cases}
 \]
  By the definition of $\eta'_{1},\eta$, for $t>0$, using
~\eqref{e.commut.eta-pr-1}, we have
 \begin{equation}\label{e.commut.tau-eta-pr-1}
   \tau(x,1,\eta'_{1}) = \chi(x, A_{1}\cup U(1,\eta))
  =\chi(x, A_{0}\cup U(1,\eta)).
 \end{equation}
  Further,
 \begin{equation}\label{e.tau.prime}
  \tau(x,t,\eta')= \begin{cases}
  \chi(x, A_{0}\cup U(0,\eta)) 				&\text{if $t=1$,}
\\ \tau(x,1,\eta') + \chi(x, U(1,\eta)\xcpt A_{0}) 	&\text{if $t=2$,}
\\ \tau(x,2,\eta')+\tau(x,t-1,\eta'_{1})-\tau(x,1,\eta'_{1}) &\text{if $t>2$.}
                   \end{cases}
 \end{equation}
  By the above definition,
 \begin{align*}
   \tau(x,1,\eta') &=\tau(x,1,\eta)+\chi(x,A_{0}),
\\ \tau(x,2,\eta') &=\tau(x,1,\eta)+\chi(x,A_{0})+\chi(x,U(1,\eta)\xcpt A_{0})
\\ & = \tau(x,1,\eta) + \chi(x,A_{0}\cup U(1,\eta))
     \ge \tau(x,1,\eta) + \chi(x, U(1,\eta))
\\ & = \tau(x,2,\eta).
 \end{align*}
  Also, from here and ~\eqref{e.commut.tau-eta-pr-1},
 \begin{equation}\label{e.tau.prime.2}
 \tau(x,2,\eta') = \tau(x,1,\eta) + \chi(x,A_{0}\cup U(1,\eta)) 
    = \tau(x,1,\eta) + \tau(x,1,\eta'_{1}).
 \end{equation}
  By domination, $\tau(x,t-1,\eta'_{1})\ge \tau(x,t-1,\eta_{1})$
and hence for all $t\ge 2$, we have, combining ~\eqref{e.tau.prime}
with ~\eqref{e.tau.prime.2},
 \begin{equation}\label{e.tau.prime.ge2}
 \begin{aligned}
  \tau(x,t,\eta') &= \tau(x,1,\eta) + \tau(x,t-1,\eta'_{1})
\\ &\ge \tau(x,1,\eta) + \tau(x,t-1,\eta_{1}) = \tau(x,t,\eta).
 \end{aligned}
 \end{equation}
\end{pproof}

 \begin{step+}{commut.main-1.equal}
  If $\tau(x,s,\eta)=\tau(x,s',\eta')$ then
$\eta(x,s)=\eta'(x,s')$.
 \end{step+}
 \begin{pproof}
  If $\tau(x,s,\eta)=0$ then clearly $\eta'(x,s)=\eta'(x,s')$ since
this means that in both processes, no progress has been made in $x$
from the initial configuration.
  Assume therefore that $\tau(x,s,\eta)>0$ and hence $s,s'>0$.

  Assume $s'=1$.
  Then $\tau(x,s,\eta)=\tau(x,1,\eta')=1$ and hence
 $x\in A_{0}\cup U(0,\eta)$.
  If $x\in U(0,\eta)$ then $s=1$ and hence the same transition that
gives $\eta'(x,1)$ also gives $\eta(x,1)$.
  Otherwise $s>1$ hence $\tau(x,s-1,\eta_{1})=1$.  
  Also, $x\in A_{0}\sbsq U(0,\eta'_{1})$, hence $\tau(x,1,\eta'_{1})=1$.
 The inductive assumption implies
$\eta'_{1}(x,1)=\eta_{1}(x,s-1)=\eta(x,s)$.
  On the other hand, ~\eqref{e.commut.eta-pr-1} and
 $x\not\in U(0,\eta)$ implies $\eta'_{1}(x,1)=\eta'(x,1)$ which
concludes this case.

  Assume now $s'>1$.
  Since $\eta(x,t)$ changes if and only if $\tau(x,t)$ does
we can assume that $x\in U(s,\eta)$ since
otherwise we can decrease $s$ without changing $\eta(x,s)$.
  The same is true for $s'$.
  Under these assumptions we have $s\ge s'$.
  By ~\eqref{e.tau.prime.ge2},
 \[
  \tau(x,s',\eta')= \tau(x,1,\eta)+\tau(x,s'-1,\eta'_{1}).
 \]
  We assumed this to be equal to
$\tau(x,s,\eta)=\tau(x,1,\eta)+\tau(x,s-1,\eta_{1})$. 
  Hence $\tau(x,s'-1,\eta'_{1})=\tau(x,s-1,\eta_{1})$.
  Also $\eta(x,s)=\eta_{1}(x,s-1)$, $\eta'(x,s')=\eta'_{1}(x,s'-1)$,
and hence the inductive assumption implies the statement. %
\end{pproof} 
\end{pproof} 

  \begin{step+}{commut.main-2}
  Let $\eta$ be a trajectory.
  Then the synchronous trajectory with initial configuration
$\eta(\cdot,0)$ dominates $\eta$.
 \end{step+}
 \begin{pproof}
  Let $A_{0}=L(0,\eta)\xcpt U(0,\eta)$.
  By ~\ref{commut.main-1} above, there is a trajectory $\eta'$ with
initial configuration $\eta(\cdot,0)$ dominating $\eta$ such that
$U(0,\eta')=U(0,\eta)\cup A_{0}=L(0,\eta)$.
  This just means that $\eta'$ is a synchronous trajectory up to time
1.
  Continuing the application of ~\ref{commut.main-1}, we can
dominate $\eta$ by a synchronous trajectory $\eta''$ up to time 2,
etc. %
\end{pproof}
  Now we can conclude the proof of the theorem as follows.
  Let $\eta$ be a trajectory with initial configuration $\xi$ and let
$\eta'$ be the synchronous trajectory with the same initial
configuration.
  Let us define
 \begin{align*}
    \sg(x,s,\xi) &= \min\setof{t:\tau(x,t,\eta')=s},
\\  \zg(x,s,\xi) &= \eta'(x,\sg(x,s)).
 \end{align*}
  To prove ~\eqref{e.t.commut}, note that due to domination,
$\tau(x,t,\eta)\le\tau(x,t,\eta')$ and hence for every $x,y,t$ there
is a $t'\le t$ with $\tau(x,t,\eta)=\tau(x,t'\eta')$.
  Let $t'$ be the first such: $t'=\sg(s,\tau(x,t,\eta))$.
  By domination, $\eta(x,t)=\eta'(x,t')=\zg(x,t)$. %
\end{Proof}

 \section{A rich example of commutative transitions}

  Let us show the known result that every transition function can be
embedded into a commutative one.
  We will use the following notation:
 \[
  b \amod m
 \]
 is the integer $x$ with $x\equiv b\pmod{m}$ and $-m/2< x \le m/2$.

 \begin{theorem}\label{t.asynch-sim}
  Let $\AA_{1}=(\CC,\SS_{1},N,\f_{1})$ be an arbitrary local (not
necessarily commutative) automaton $N(x)$.
  Then there is an automaton $\AA_{2}=(\CC,\SS_{1}\times R,N,\f_{2})$,
where for $s\in \SS_{1}\times R$ we write $s=(s.\F,s.\G)$, with the
following property.

  Let $\xi_{1}$ be an arbitrary configuration of $\f_{1}$ and let $\xi_{2}$
be a configuration of $\f_{2}$ such that for all $x$ we have
$\xi_{2}(x)=(\xi_{1}(x), 0\cdots 0)$.
  Then for the synchronous trajectory $\eta_{2}$ of $\f_{2}$, with initial
configuration $\xi_{2}$, the space-time configuration $(x,t)\mapsto
\eta_{2}(x,t).\F$ is a synchronous trajectory of $\f_{1}$.
  Moreover, in this trajectory, the state of each cell changes in each
step.
 \end{theorem}

  In other words, as long as we update synchronously the rule $\f_{2}$
behaves in its field $\F$ just like the arbitrary rule $\f_{1}$.
  But $\f_{2}$ has invariant histories, so it is much more robust.

\begin{proof}
  Let $\SS_{2} = \SS_{1}^{2}\times\{0,1,2\}$.
  The three components of each state $s$ of $\SS_{2}$ will be written
as 
 \[
  s.\Cur, s.\Prev\in\SS_{1},\; s.\Age\in\{0,1,2\}.
 \]
  The statement of the theorem will obtain by $s.\F=s.\Cur$,
$s.\G=(s.\Prev,s.\Age)$.
  The field $\Age\in\{0,1,2\}$ will be used to keep track of the time of
the simulated cells mod 3, while $\Prev$ holds the value of $\Cur$ for the
previous value of $\Age$.

  Let us define $s' = \f_{2}(\xi)(x)$.
  If there is a $y\in N(x)$ such that
 $(\xi(y).\Age-\xi(x).\Age) \amod 3 < 0$ (i.e.~some neighbor lags
behind) then $s'=\xi(x)$ i.e.~there is no effect.
  Otherwise, let $\sg(y)$ be $\xi(y).\Cur$ if
$\xi(y).\Age=\xi(x).\Age$, and $\xi(y).\Prev$ otherwise.
 \begin{align*}
     s'.\Cur  &= \f_{1}(\sg)(x),
 \\  s'.\Prev &= \xi(x).\Cur,
 \\  s'.\Age  &= \xi(x).\Age + 1 \bmod 3.
 \end{align*}
  Thus, we use the $\Cur$ and $\Prev$ fields of the neighbors
according to their meaning and update the three fields according to
their meaning.
  It is easy to check that this transition rule simulates $\f_{1}$
in the $\Cur$ field if we start it by putting 0 into all other fields.

  Let us check that $\f_{2}$ is locally commutative.
  If two neighbors $x,y$ are both are allowed to update then neither
of them is behind the other modulo 3, hence they both have the same
$\Age$ field.
  Suppose that $x$ updates before $y$.
  In this case, $x$ will use the the $\Cur$ field of $y$ for
updating and put its own $\Cur$ field into $\Prev$.
  Next, since now $x$ is ``ahead'' according to $\Age$, cell $y$ will
use the $\Prev$ field of $x$ for updating: this was the $\Cur$ field
of before.
  Therefore the effect of consecutive updating is the same as that of
simultaneous updating.
\end{proof} 

  The commutative medium of the above proof is also called the ``marching
soldiers'' scheme since its handling of the $\Age$ field reminds one of a
chain of soldiers marching ahead in which two neighbors do not want to be
separated by more than one step.
  It is shown in ~\cite{BermSim88} that if the update times obey a Poisson
process then the average computation time of this simulation within a
constant factor of the computation time of the synchronous computation.

\begin{remark}
  In typical cases of asynchronous computation, there are more
efficient ways to build a commutative rule than to store the whole
previous state in the $\Prev$ field.
  Indeed, the transition function typically does not use the complete
state of cells in $N(x)$.
  Rather, the cells only ``communicate'' in the sense that there is a
message field and the next state of $x$ depends only on this field of
the neighbor cells.
  In such cases, it is sufficient in the above construction to store
the previous value of this message field.
  We can sometimes decrease the message field by taking several steps
of $\f_{2}$ to simulate a single step of $\f_{1}$.
  \end{remark}

  In case of one-dimensional systems, the ``marching soldiers'' scheme has
the following strengthening of the original property saying that $\zg(x,t)$
is independent of the order of updating.
as in Example ~\ref{x.CA}

  \begin{theorem}\label{t.1dim-prop}
  Let $\AA_{1}=(\CC,\SS_{1},N,\f_{1})$ be an arbitrary one-dimensional
cellular automaton defined, as in Example ~\ref{x.CA}, via a transition
function $g$.
  Let the automaton $\AA_{2}=(\CC,\SS_{1}\times R,N,\f_{2})$
be defined as in the proof of Theorem ~\ref{t.asynch-sim}.
  Let $\eta$ be an arbitrary asynchronous trajectory of $\AA_{2}$.
  Let us define the functions $\dg(x)$, $\bar\eta(x,u)$ by $\dg(0)=0$, and
 \begin{align*}
    \dg(x+1)   &= \dg(x)+\eta(x+1,0).\Age-\eta(x,0).\Age,
\\  \bar\tau(x,t) &= \tau(x,t)+\dg(x),
\\  \bar\eta(x,u) &= \zg(x, u - \dg(x)).\Cur
 \end{align*}
  for all $u$ of the form $\bar\tau(x,t)$.
  Also, let $\bar\eta(x,\dg(x)-1)=\eta(x,0).\Prev$.
  Then $\tau(x,t)>0$ implies with $u=\bar\tau(x,t)-1$ that 
 \[
 \bar\eta(x,u+1) = g(\bar\eta(x-1,u),\bar\eta(x,u), \bar\eta(x+1,u))
 \]
  and all terms in this equation are defined.
  \end{theorem}

  The proof is straightforward verification.
  The theorem essentially says that from each asynchronous trajectory
$\eta$ of $\AA_{2}$, some synchronous trajectory $\bar\eta$ of $\AA_{1}$ can
be reconstructed as $\bar\eta(x,u)=\zg(x,u-\dg(x)).\Cur$.
  The function $\dg(x)$ shows how much ``ahead'' or ``behind'' we are in
simulating this trajectory when we start in $\eta$.

  \begin{remark}
  This theorem fails in other neighborhood structures, namely in networks
containing cycles: there, only certain initial configurations
$\eta(\cdot,0)$ allow the construction of $\dg(x)$.
  In the ones that do not allow it, there is some inconsistency in the
timing function $\eta(x,0).\Age$ (a loop along which the sum of local
increments of $\Age$ is not 0).
  In a connected network, this loop will imply that each cell can have only
finitely many state changes, even in an infinite trajectory.
 \end{remark}

\section{Undecidability}

\begin{lemma}\label{l.turing}
  Let us be given a one-dimensional commutative cellular automaton
over the set of natural numbers, with ``free boundary condition'', by a set
of states $\SS=\{0,\ldots,n-1\}$, transition functions $g:\SS^{3}\to \SS$
and $g_{0}:\SS^{2}\to \SS$ as in Example ~\ref{x.CA}, with
 $g(0,0,0)=0$, $g_{0}(1,s)=1$ (for all $s$).
  The following problem is undecidable, as a function of $n,g,g_{0}$:
  Is there any synchronous trajectory of this cellular automaton, with
$\eta(x,0)=0$ for all $x$ and $\eta(0,t)=1$ for some $t>0$?
  \end{lemma}

  \begin{proof}
  There is a standard construction to simulate Turing machines with such
cellular automata, so the question reduces to the question whether an
arbitrary Turing machine will halt when started on an empty tape.
  \end{proof}

  \begin{lemma}\label{l.undec}
  Let us be given a one-dimensional commutative cellular automaton over the
set of natural numbers, with ``free boundary condition'', by a set of
states $\SS=\{0,\ldots,n-1\}$, transition functions $g:\SS^{3}\to \SS$
and $g_{0}:\SS^{2}\to \SS$ as in Example ~\ref{x.CA}.
  
  The following problem is undecidable, as a function of $n,g,g_{0}$:
  Is there any trajectory of this cellular automaton, with $\eta(0,0)=0$
and $\eta(0,t)=1$ for some $t>0$?
  \end{lemma}

   Of course, once the automaton is commutative it does not matter whether
the trajectory asked for is synchronous or asynchronous.

  \begin{proof}
  From now on, without danger of confusion, let us write
$g(r,s)=g_{0}(r,s)$ and forget about $g_{0}$.
  Let us be given a cellular automaton $g$ like in Lemma
~\ref{l.turing}, with state set $\SS=\{0,\ldots,n-1\}$.
  We construct a new cellular automaton over the set of states
$\SS'=\SS\cup\{n\}$, with the following transition function $g'$.
  Over states $s<n$, the functions $g'$ behave as $g$.
  Further, we have the following rules for $g'$ when at least one of the
arguments is $n$.
 \begin{align*}
      (n,s) &\to g(0,0),
\\    (s,n) &\to g(s,0) &&\text{ for } s < n,
\\    (n,r,s) &\to n,
\\    (r,n,s) &\to g(r,0,0) &&\text{ for } r < n,
\\    (r,s,n) &\to g(r,s,0) &&\text{ for } r,s < n,
 \end{align*}
  and $(r,s,n)\to s$, $(r,s)\to r$ in all remaining cases.
  By these rules, the symbol $n$ ``sweeps'' right and in its wake the rule
$g$ will operate as if it had started from the a configuration of all 0's.
  Thus, let $\eta$ be the synchronous trajectory of $g$ with $\eta(x,0)=0$
for all $x$.
  Then clearly if $\eta'$ is any synchronous trajectory of $g'$ with
$\eta'(0,0)=n$ then for all $t>0$, for all $x\le t$ we have
$\eta'(x,t)=\eta(x,t)$.
  
  Let us now apply the construction of the proof of Theorem
~\ref{t.asynch-sim} to $g'$ to obtain commutative rule $g''$ over the set
of states $\SS''=(\SS')^{2}\times\{0,1,2\}$.
  We will prove that $g''$ has an asynchronous trajectory $\eta''$ with
$\eta''(0,0)=(n,0,0)$ and $\eta''(0,u)=(1,1,0)$ for some $u$, if and only
if $g$ has a synchronous trajectory $\eta$ with $\eta(0,x)=0$ for all $x$
and $\eta(0,u)=1$ for some $u$.
  Since we know that the question whether this happens is undecidable from
$g$, we will have proved that the question whether some cellular automaton
has an asynchronous trajectory $\eta$ with $\eta(0,0)=s_{0}$ and
$\eta(0,u)=s_{1}$ for some $s_{0}\ne s_{1}$ is undecidable; this will
complete the proof.

  The ``if'' part:
  Suppose first that $g$ has a synchrounous trajectory $\eta$ with
$\eta(0,x)=0$ for all $x$, and and $\eta(0,u)=1$ for some $u$.
  As mentioned above, then the synchronous trajectory $\eta'$ of $g'$ has
$\eta'(x,t)=\eta(x,t)$ for all $x\le t$.
  Consider the synchronous trajectory $\eta''$ of $g''$ started from
$\eta''(x,0)=(n,0,0)$ for all $x$.
  Then for all $t>0$ and all $x\le t$ we have
 \[
  \eta''(x,t ) = (\eta'(x,t),\eta'(x,t-1),t\bmod 3) =
  (\eta(x,t),\eta(x,t-1),t\bmod 3).
 \]
  Let $v$ be the first number $>u+1$ divisible by 3.
  We have
 \[
  \eta''(0,v) = (\eta(0,v),\eta(0,v-1),0) = (1,1,0).
 \]
  The ``only if'' part:
  Assume that $\eta''$ is an asynchronous trajectory of $g''$ with
$\eta''(0,0)=(n,0,0)$ and $\eta''(0,w)=(1,1,0)$ for some $w$.
  Then $\tau''(0,w)>0$ and defining $u=\bar\tau''(0,w)-1$, Theorem
~\ref{t.1dim-prop} implies  
 \[
 \bar\eta''(0,u+1) = g'(\bar\eta''(0,u), \bar\eta''(1,u)).
 \]
  Applying theorem  repeatedly, we obtain
 \[
 \bar\eta''(0,v+1) = g'(\bar\eta''(x-1,v),\bar\eta''(x,v),
\bar\eta''(x+1,v))
 \]
  or, if $x=0$, the same relation with the first argument of $g'$ omitted,
for $v=0,\ldots,u$ and $x\le \min\{v,(u-v)\}$.
  Now, if $\eta''(0,w)=(1,1,0)$ then $\bar\eta''(0,u+1)=1$ while
$\bar\eta''(0,0)=n$.
  We have just found that $\bar\eta''(x,v)$ develops according to $g'$
for $v=0,\ldots,u$ and $x\le \min\{v,(u-v)\}$.
  As discussed above, therefore $\bar\eta''(0,u+1)=1$ if and only if $g$
computes 1 at $(0,u+1)$ from an all-0 initial configuration.
  \end{proof}
  
 \begin{proof}[Proof of Theorem~\protect\ref{t.undec}]
  Let the local state space be the set of integers
$\SS=\{0,\ldots,n+2\}$.
  Let $g:\SS_{0}^{3}\to \SS_{0}$ and $g_{0}:\SS_{0}^{2}\to \SS_{0}$ be the
rules for a commutative cellular automaton transition rule with state set
$\SS_{0}=\{0,\ldots,n-1\}$.
  We define the transition function $\f$.
  We will write $\f(x,y,z)=y'$ as $(x,y,z)\to y'$.
  We require
 \begin{align}
\label{e.branch-1}
   (s, n, 0)   &\to n+1,
\\\label{e.branch-2}
   (s, n, 1)   &\to n+2,
\\\label{e.CA-traj-0}
   (r, s, t) &\to g_{0}(s,t), &&\text{ for all } r\ge n,\ r,s < n,\ r\ne 1,
\\\label{e.CA-traj-1}
   (r, s, t)   &\to g(r, s, t) &&\text{ for all } r, s, t < n,
\\\label{e.CA-traj-2}
   (r, s, t)   &\to g(r, s, 0) &&\text{ for all } r, s < n,\ t\ge n,
 \end{align}
 and $(r, s, t)\to s$ in all remaining cases.
  Let us show that $\f$ has invariant histories if and only if $g$ has no
asynchronous trajectory $\eta_{0}$ over $\CC=\ZZ_{+}$ with
$\eta_{0}(0,0)=0$ and $\eta_{0}(0,t)=1$ for some $t$.
  Assume first that $g$ has such a trajectory.
  Let us define the initial configuration $\xi$ of $\f$ as
$\xi(x)=n$ if $x=-1$ and 0 otherwise.
  We may apply rule ~\eqref{e.branch-1} first to get $\eta(-1,1)=n+1$.
  Or, we may apply rules
~\eqref{e.CA-traj-0},\eqref{e.CA-traj-1},\eqref{e.CA-traj-2} first to cells
$x>0$ on the right repeatedly.
  Sooner or later we have $\eta(0,t)=1$, which allows $\eta(-1,t+1)=n+2$ by
rule ~\eqref{e.branch-2} in the next step.
  Thus, depending on the order of rule application, we obtained in cell
$-1$ the sequence $n,n+1$ or $n,n+2$.

  Suppose now that $g$ has no such trajectory and let $\xi$ be an arbitrary
configuration of $\f$.
  Each occurrence of a state $\ge n$ remains such an occurrence.
  On segments between them, the commutative rule $g$ works.
  The only other transitions possible are $(r,n,0)\to n+1$ and
$(r,n,1)\to n+2$.
  Assume $\eta(x,0)=n$ and consider the sequence
of different values in $\eta(x+1,t)$.
  Let us show that 0 and 1 cannot both occur in this sequence and hence
only one of the transitions is possible.
  Indeed, if 0 occurs before 1 then our assumption about $g$ excludes the
occurrence of 1 in the sequence any later.
  If 1 occurs in the sequence before 0 then our rules (in particular,
~\eqref{e.CA-traj-0}) do not allow any change of the state of $x+1$ after
that.
 \end{proof}

\section*{Acknowledgments}
  
  I thank Robert Solovay for pointing out several errors in the first
version, Wayne Snyder for calling my attention to ~\cite{HuetConflu80} and
the anonymous referee for his careful reading and many corrections.
  

\end{document}